\documentclass[aps,twocolumn,showpacs,preprintnumbers,pra]{revtex4}
\usepackage{graphicx}
\usepackage{dcolumn}
\begin{document}                
\author{J.~Ries}
\author{B.~Brezger}
\author{A.~I.~Lvovsky}
\address{Fachbereich Physik, Universit\"at Konstanz, D-78457
Konstanz, Germany}
\homepage{http://www.uni-konstanz.de/quantum-optics/quantech/}

\date{\today}
\title{Experimental Vacuum Squeezing in Rubidium Vapor via Self-Rotation}%

%
\hyphenation{cryp-to-gra-phy}
\newcommand{\Name}[1]{#1,}
\newcommand{\Vol}[1]{{\bf #1,}}
\newcommand{\Year}[1]{(#1)}
\newcommand{\Page}[1]{#1}
\newcommand{\Book}[1]{{\it #1}}
\newcommand{\Publ}[1]{(#1)}
\newcommand{\Review}[1]{{#1}}
\newcommand{\And}[1]{and #1}
\begin{abstract}
We report the generation of optical squeezed vacuum states by
means of polarization self-rotation in rubidium vapor following a
proposal by Matsko {\em et al.} [Phys.\ Rev.\ A {\bf 66}, 043815
(2002)]. The experimental setup, involving in essence just a diode
laser and a heated rubidium gas cell, is simple and easily
scalable. A squeezing of $(0.85\pm0.05)$ dB was achieved.
\end{abstract}

\pacs{42.50.Lc, 42.50.Dv, 42.50.Nn}
\maketitle

\paragraph{Introduction}
Squeezed states of a radiation field are nonclassical states whose
field-quadrature noise is phase dependent with a minimum below the
shot noise level. First investigated theoretically \cite{Walls83}
and generated experimentally \cite{Slusher85,Wu86} in the 1980's,
these states awoke initial interest as a way of overcoming the
shot-noise precision restrictions in optical measurements
\cite{Grangier87,Xiao87} and enhancing the capacity of
communication channels \cite{YuenShapiro80}. Further interest to
squeezed light was triggered by recent advances in the field of
quantum information and communication. In this context they are
viewed primarily as a means of generating quadrature entanglement
\cite{Ou92}, a valuable resource that can be applied, for example,
for quantum teleportation \cite{Furusawa98}, computation
\cite{LLoyd99}, cryptography \cite{Ralph00} and manipulation of
atomic quantum states \cite{Kuzmich00}. Protocols of error
correction \cite{Braunstein982,Lloyd98} and entanglement
purification \cite{Duan00} have been elaborated, establishing
continuous variable entanglement as one of the most promising
tools for quantum information technology.


A number of techniques for producing squeezed states of light have
been developed, using for instance four wave mixing
 in atomic vapor \cite{Maeda87}, Kerr effect in cold atoms
\cite{Lambrecht96} or optical fibers \cite{Shelby86,Silberhorn01},
optical parametric amplifiers \cite{Wu86}. All these methods
require rather complicated setups that cannot be scaled easily,
effectively preventing their practical application for the quantum
information purposes. In the present paper we demonstrate a
simple, scalable squeezed vacuum source which consists in essence
of a continuous-wave (cw) diode laser and an atomic rubidium vapor
cell. Easy and inexpensive generation of multiple
Einstein-Podolsky-Rosen-entangled modes lies at hand, making them
available to a variety of experiments and applications.

We make use of a phenomenon known as cross-phase modulation
squeezing \cite{Haus95}: if linearly polarized light propagates
through a nonlinear medium in which elliptical light would undergo
polarization self-rotation, the orthogonal vacuum mode gets
squeezed. Self-rotation and the associated squeezing effect are
present in any Kerr medium but are especially strong in atomic
vapor due to its large optical nonlinearity near resonance. To
achieve squeezing it is sufficient to use a moderately intense cw
diode laser as a pump. Pulsed lasers or cavities are not needed
and the diode laser does not even have to be frequency stabilized.

\paragraph{Theory}
The theory of squeezing via polarization self-rotation for the
single mode case has been given in Ref.~\cite{Matsko02}. Here we
present its brief quasiclassical overview. A detailed multimode
description including the degrading effects of resonance
fluorescence, phase mismatch and absorption will be published
elsewhere.

Interaction with an elliptically polarized light field can cause
an initially isotropic  $\chi^{(3)}$-nonlinear medium to become
circularly birefringent: the two circular components of different
intensities propagate with different phase velocities.
Specifically in atomic vapor, the mechanisms leading to this
birefringence include optical pumping, ac Stark shifts, as well as
ac Stark shift induced quantum beats and Bennett structures
\cite{Rochester01}. Propagating through a cell of length $l$, the
polarization ellipse of light will rotate by an angle $\varphi$
which is proportional to the ellipticity $\epsilon\ll 1$:
\begin{equation}
    \varphi=g l\epsilon,
\end{equation}
with the self-rotation parameter $g$ dependent on the incident
light intensity and frequency.

\begin{figure}[tbp]
\begin{center}
\includegraphics[width=0.9\columnwidth]{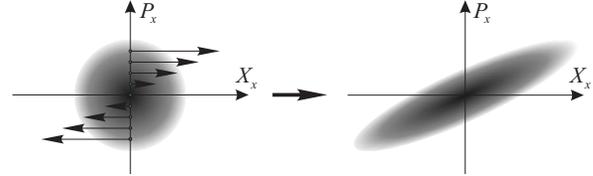}
\caption{\label{squeezephasespace}Transformation of the phase
space: a Wigner function describing the state of the probe field
in the mode polarized orthogonally to the strong pump beam.
Self-rotation results in a linear shear of phase space
(Eqs.~\ref{classquadsqueeze} and~\ref{classPconserve}). The
circular phase-space quasiprobability density describing the
vacuum state is transformed into squeezed vacuum.}
\end{center}
\end{figure}

Consider a monochromatic, elliptically polarized light field of
frequency $\omega$ propagating in the $z$-direction. The complex
field amplitudes in the two linear polarization components are
given by $\mathcal{E}_x$ and $\mathcal{E}_y$, with
\mbox{$|\mathcal{E}_x|\ll |\mathcal{E}_y|$.} Assuming the ``pump"
amplitude $\mathcal{E}_y$ real, we decompose the ``probe"
$x$-field into two real quadrature components in and out of phase
with the pump field: \mbox{$\mathcal{E}_x(z)=X_x(z)+i P_x(z)$}.
Then the ellipticity $\epsilon$ of the laser light is given by
\begin{equation}
   \epsilon\approx\frac{P_x(z)}{\mathcal{E}_y(z)}. \label{ellipticity}
\end{equation}

The resulting self-rotation by $\varphi\ll 1$ causes the pump
field to contribute its fraction $\varphi\mathcal{E}_y$ into the
$x$-polarization, namely into the $X_x$ quadrature which is in
phase with $\mathcal{E}_y$:
\begin{eqnarray}
 X_x(l) &=& X_x(0) + \varphi \mathcal{E}_y = X_x(0) + g l P_x(0) \label{classquadsqueeze}\\
 P_x(l) &=& P_x(0), \label{classPconserve}
\end{eqnarray}
$\mathcal{E}_y$ and $\epsilon$ remain effectively unchanged.

We now assume the $x$-polarized light field as classical noise
described by a probability distribution in the quadrature plane
(Fig.~1). To describe slow fluctuations, we regard the propagation
of a light field described by $X_x(z),\ P_x(z)$ and
$\mathcal{E}_y$ through the medium under stationary conditions. A
random quadrature component $P_x(0)$ results in a small random
ellipticity according to Eq.~(\ref{ellipticity}). The propagation
through the cell will displace the point ($X_x,\ P_x$) in the
phase space as described in Eqs.~(\ref{classquadsqueeze}) and
(\ref{classPconserve}). The phase space probability distribution
experiences a linear shear, resulting in reduction of the noise in
some phase quadratures below the original level.

Although the above treatment is purely classical, it is completely
identical in application to the quantum quadrature noise.
Considering the Heisenberg evolution of the field-quadrature
operators $\hat X_x(z)$ and $\hat P_x(z)$ we notice that the
evolution described by Eqs.~(\ref{classquadsqueeze}) and
(\ref{classPconserve}) can be associated with the interaction
Hamiltonian $\hat H=\hbar g c \hat P_x^2/2$, $c$ being the speed
of light in the medium. Under a second order Hamiltonian, each
point of the Wigner function
--- the quantum-mechanical analogue of the phase-space probability
distribution --- evolves according to the classical equations of
motion \cite{Schleich}. The incoming symmetric Gaussian Wigner
function of the vacuum is therefore transformed into an elliptical
Wigner function of a minimum uncertainty squeezed vacuum state.

In a homodyne measurement such a state results in a quadrature
noise depending on the local oscillator phase $\chi$
(cf.~\cite{Matsko02}):
\begin{eqnarray}
\label{quadnoise}
 \lefteqn{\langle\Delta E_x(\chi,l)^2\rangle}\\
 && = \frac{\mathcal{E}_0^2}{4}
  \left((1-\alpha l)(1-2gl\sin\chi\cos\chi+g^2 l^2\cos^2\chi)+\alpha
  l\right).\nonumber
\end{eqnarray}
Here $\frac{\mathcal{E}_0^2}{4}$ is the standard quantum noise
limit (SQL) associated with the vacuum state, and we have taken
into account absorption of a small fraction $\alpha l\ll 1$ of the
squeezed light after passing the medium. This treatment of
absorption is approximate because some of the light energy is
actually lost in the atomic vapor along the propagation path.
However, it permits to account for one essential feature of our
experimental result: the minimum uncertainty property is lost when
$g l>0$ and $\alpha l>0$.

\paragraph{Experimental apparatus}
\begin{figure}[tbp]
\begin{center}
\includegraphics[width=1\columnwidth]{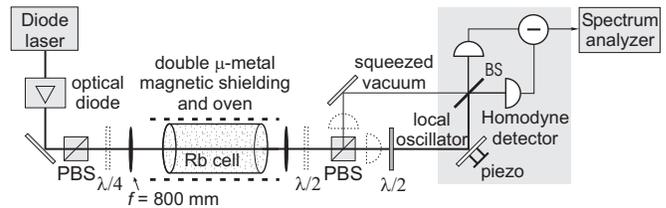}
\caption{\label{setup} Experimental setup. The second polarizing
beam splitter (PBS) separates the $x$-polarized mode containing
the squeezed vacuum from the $y$-polarized pump beam. The
$\lambda/2$ plate rotates the polarization of the pump which is
then used as the local oscillator in balanced homodyne detection,
the piezoelectric transducer scans its phase. Dashed contours show
optical elements used in the self-rotation measurement and/or the
mode-matching alignment (see text). }
\end{center}
\end{figure}
The scheme of our experimental setup is shown in Fig.~\ref{setup}.
A free running external cavity diode laser (Toptica DL-100)
delivered 35\,mW of optical power in a single mode at a wavelength
of 780\,nm, tunable across the rubidium $D2$ line. The beam was
shaped with a pair of cylindrical lenses and a polarizing beam
splitter (PBS) purified the linear polarization. The beam was
focused by a lens ($f=800\,$mm) into a heated and magnetically
shielded $75\,$mm rubidium cell which had optical quality
anti-reflection coated windows and was filled with the nearly pure
isotope $^{87}$Rb. With a focal beam diameter of 600 $\mu$m we
operated in a high saturation regime.

To determine the polarization self-rotation parameter $g l$ a
fine-adjustable $\lambda/4$ plate was placed in front of the cell.
Its rotation angle in radians was equal to the generated
ellipticity $\epsilon$. The outputs of the second PBS were
directly fed into two photodiodes. A $\lambda/2$ plate placed
after the cell was adjusted so that the DC photodiode signals
$S_1$ and $S_2$ were exactly equal when the laser was detuned far
from the resonance. This ``balanced polarimeter'' was sensitive to
the polarization ellipse rotation while being insensitive to
changes in ellipticity. The rotation angle $\varphi \ll 1$ can be
determined from the photodiode signal
difference~\cite{Rochester01}:
\begin{equation}
\varphi = \frac{S_1-S_s}{2(S_1+S_2)}
\end{equation}

\begin{figure}[tbp]
\begin{center}
\includegraphics[width=\columnwidth]{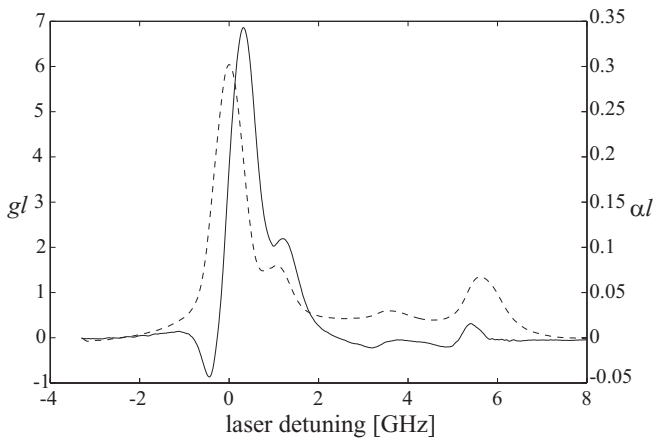}

\caption{\label{alphagl} Polarization self-rotation parameter $gl$
(solid line) and absorption coefficient $\alpha l$ (dashed line),
measured across the $^{87}$Rb $D2$ line, for rubidium vapor at a
temperature of $70^\circ$C. The two main features correspond to
the hyperfine levels of the $^{87}$Rb ground state. The hyperfine
structure of the excited state is not resolved because it is
narrower than both Doppler and power broadening. The ``side
lobes'' around 1 and 4\,GHz are attributed to some content of the
$^{85}$Rb isotope. Zero detuning corresponds to the maximum
absorption point.}
\end{center}
\end{figure}

By recording  $S_1$ and $S_2$ while scanning over the rubidium
$D2$ line for many different initial small ellipticities
$\epsilon\ll 1$, the self-rotation parameter $g l$ for each laser
detuning was determined by polynomial fitting. Fig.~\ref{alphagl}
shows $g l$ and the absorption coefficient $\alpha l$ for our
working temperature of $70^\circ$C. Note that a relatively high
transmission on resonance is due to very high saturation of the
atomic transition
--- in the linear small signal regime, the rubidium vapor is
optically dense.

To measure the quantum field quadrature noise we employed the
standard balanced homodyne detection technique. The $y$-polarized
component of the laser beam emerging from the cell was used as the
local oscillator. To this end it was first separated from the
$x$-polarized probe mode by means of a polarizing beam splitter
and its polarization was rotated by $90^\circ$. It was then
overlapped with the squeezed mode on a symmetric beam splitter. To
achieve good mode matching, an auxiliary $\lambda/2$ plate placed
directly after the cell was set to provide equal splitting into
both polarization modes. The resulting Mach-Zehnder interferometer
was carefully aligned for a maximum visibility of $V=0.98\pm 0.01$
corresponding to a mode matching efficiency of $V^2\approx 0.96$.
The auxiliary $\lambda/2$ plate was then removed.

A piezoelectric element was used to scan the local oscillator
phase and thus the field quadrature whose fluctuations were
measured. The intensities at the beam splitter outputs were
measured with Si PIN photodiodes (Hamamatsu S3883 with a nominal
quantum efficiency of 91\%). The AC signals of the diodes were
amplified with a low noise amplifier with a cutoff frequency of
approximately 50 MHz and subtracted by a hybrid junction (H-9 from
M/A-COM). The noise was measured with a spectrum analyzer operated
in the zero span mode at a set of radio frequencies $\Omega_{RF}$
between 3 and 30 MHz with a resolution bandwidth of 1 MHz.

\paragraph{Results and discussion}
\begin{figure}[tbp]
\begin{center}
\includegraphics[width=\columnwidth]{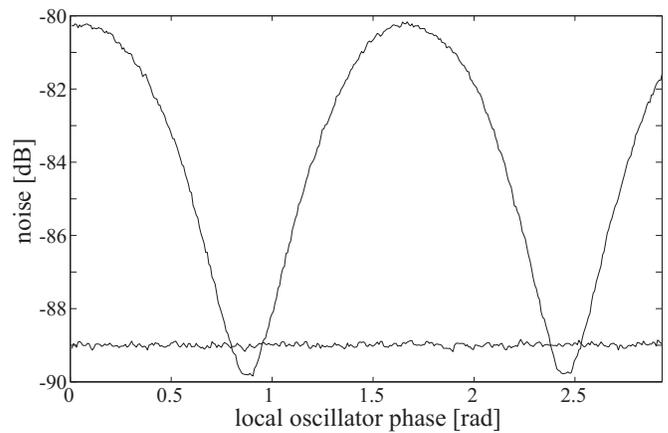}
\caption{\label{squeezedb} Phase dependent noise and shot noise
(measured by blocking the squeezed vacuum mode). Measured at
$\Omega_{RF}= 5$ MHz with a laser detuning of $0.35\,$GHz, a
resolution bandwidth of 1 MHz, a video bandwidth of 100 Hz, and a
cell temperature of $70^\circ$C. A squeezing of 0.85 dB was
achieved.}
\end{center}
\end{figure}
The squeezing homodyne measurements were performed at a finite
radio-frequency $\Omega_{RF}$, as usual, because of the high
optical and electronic noise near DC. Fig.~\ref{squeezedb} shows
the phase dependent noise of the $x$-polarization mode which
contains the squeezed vacuum. It is compared to the SQL which can
be measured by blocking the squeezed vacuum path. The phase
dependent noise minima fall below the SQL exhibiting a squeezing
of $(0.85 \pm 0.05)\,$dB. This corresponds to a squeezing of
$(1.23 \pm 0.07)\,$dB at the cell output when corrections for
linear losses, inefficient photodiodes and electronic
photodetector noise are made. The geometric mean noise of the
squeezed mode, however, strongly exceeds the SQL so the observed
ensemble is not a minimum uncertainty state. The excess noise
cannot be explained by absorption alone and is presumably due to
resonance fluorescence into the optical mode probed.

\begin{figure*}[tbp]
\begin{minipage}[t]{\columnwidth}
\includegraphics[width=\columnwidth]{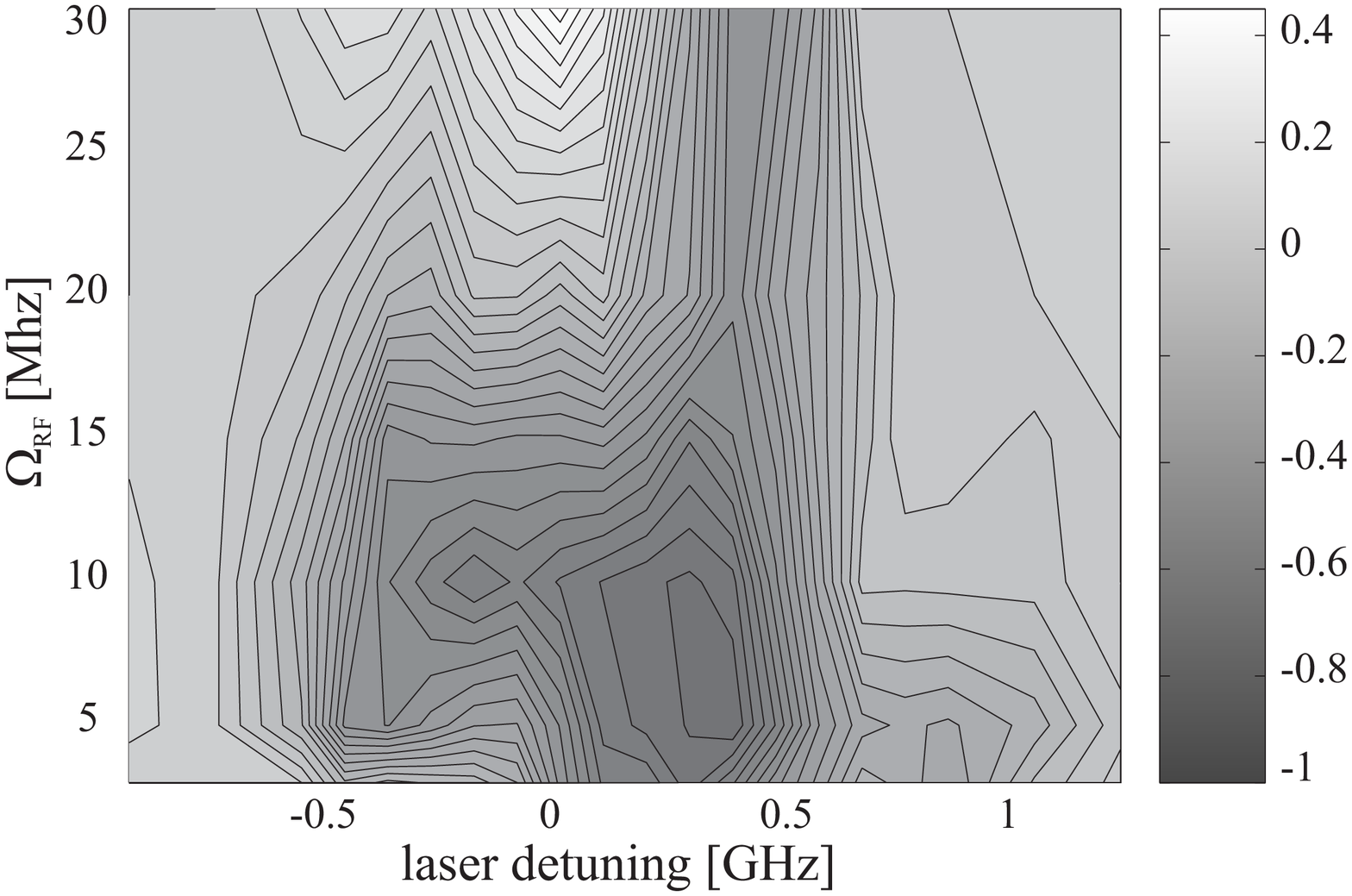}
\caption{\label{squeezenuom} Minimum quadrature noise level
relative to the SQL (in dB) for a range of radio frequencies
$\Omega_{RF}$ and laser detunings across one hyperfine component
of the rubidium D2 line. The electronic noise has been
subtracted.}
\end{minipage}
\hfill
\begin{minipage}[t]{\columnwidth}
\includegraphics[width=\columnwidth]{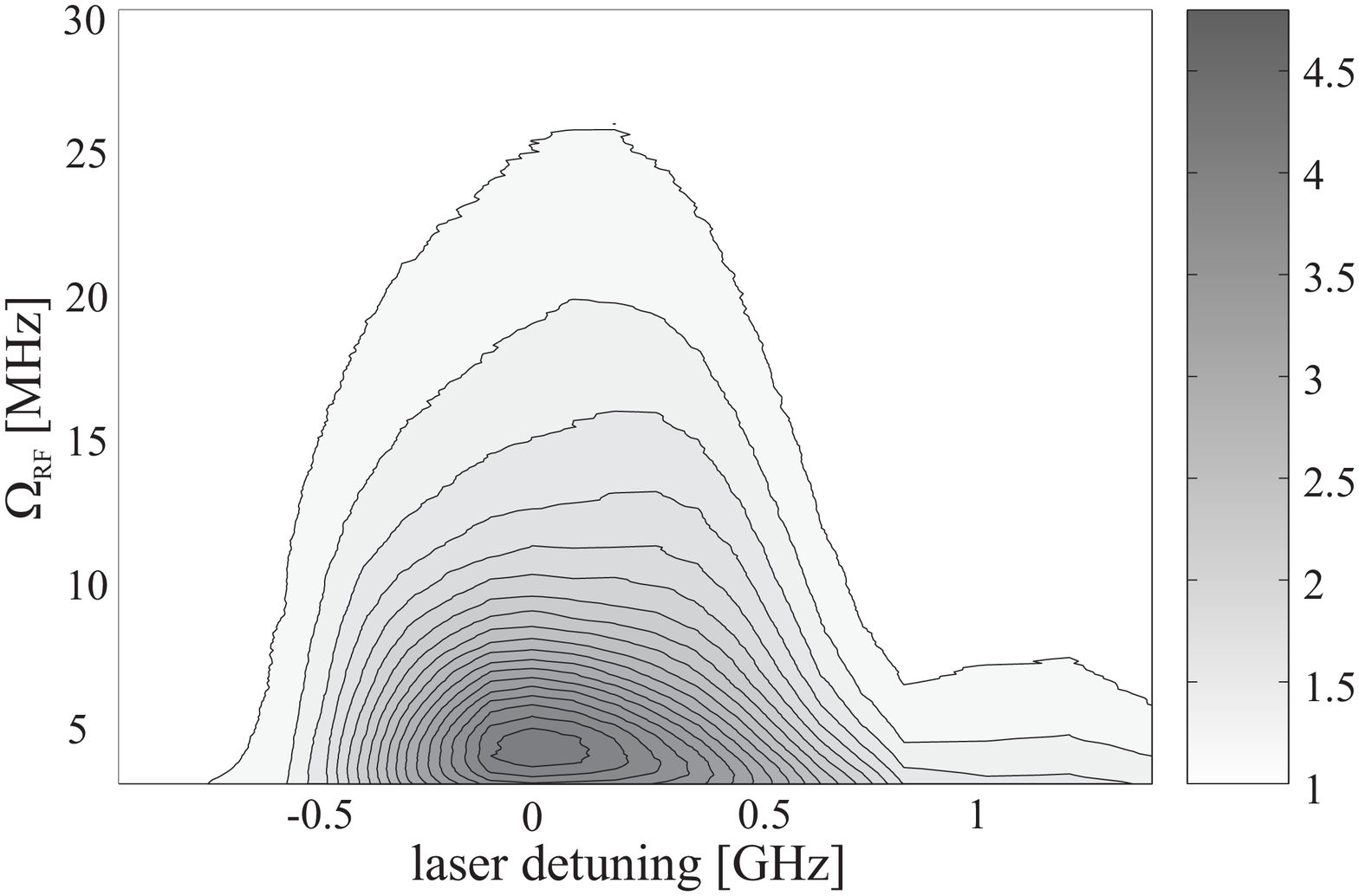}
\caption{\label{noise} Excess noise due to resonance fluorescence
on a linear scale, in units of the vacuum noise for a range of
radio frequencies $\Omega_{RF}$ and laser detunings. Measured by
reducing the mode matching while keeping a good spatial overlap of
the pump and the probe mode in the vapor cell.}
\end{minipage}
\end{figure*}

Fig.~\ref{squeezenuom} shows the squeezing as a function of the
laser detuning and the radio frequency $\Omega_{RF}$. The
variation of squeezing with the laser detuning follows
approximately the self-rotation curve in Fig.~\ref{alphagl}. For
small $\Omega_{RF}$ the squeezing is reduced due to extra noise
from resonance fluorescence which scatters light into our vacuum
mode. This extra noise was measured by misaligning the mode
matching until the phase dependence vanished and maximizing the
spatial overlap between the pump and probe modes in the cell by
maximizing the noise. Fig.~\ref{noise} shows the extra noise
normalized to the vacuum noise. Its spectral width approximates
the natural linewidth of $\Gamma=6\,$MHz
\cite{resonancefluorescence}, its dependence on the laser detuning
is similar to the absorption line (Fig.~\ref{alphagl}). For
$\Omega_{RF}\gg\Gamma$ the excess noise becomes negligible.
Maximum squeezing is achieved for $\Omega_{RF}\sim 5\,$MHz to
$10\,$MHz. The reason for the squeezing to degrade with higher
$\Omega_{RF}$ might be a growing phase mismatch between the pump
and the probe expected in the neighborhood of an atomic resonance.

\paragraph{Summary and outlook}
We reported the generation of squeezed optical vacuum states in
rubidium vapor with a simple setup. A squeezing of 0.85\,dB was
achieved which is among the best atomic vapor squeezing results
accomplished. The squeezing of 6 to 8 dB predicted in
\cite{Matsko02} was not reached due to the degrading effects of
self-focusing, phase mismatch and resonance fluorescence. Further
theoretical investigation of these effects will help us optimize
the experimental parameters. The self-rotation parameter and the
associated squeezing might increase by using a cell with a
moderate amount of buffer gas \cite{Novikova02}. We are optimistic
that a noticeable improvement of the squeezing is possible. By
splitting the laser and directing it into several rubidium cells,
numerous squeezed states, which are phase locked with respect to
each other, can be generated easily.

\paragraph{Acknowledgement}
We acknowledge helpful discussions with K.~P.~Marzlin,
J.~H.~Shapiro and I.~Novikova, as well as financial support from
the Deutsche Forschungsgemeinschaft and the Optik-Zentrum
Konstanz.


\end{document}